\input epsf
\input harvmac

%%%%%%%%%%%%%%%%%%
%
%  This inputs the macro package epsf.tex
%
\ifx\epsfbox\UnDeFiNeD\message{(NO epsf.tex, FIGURES WILL BE IGNORED)}
\def\figin#1{\vskip2in}% blank space instead
\else\message{(FIGURES WILL BE INCLUDED)}\def\figin#1{#1}\fi
\def\ifig#1#2#3{\xdef#1{fig.~\the\figno}
\goodbreak\midinsert\figin{\centerline{#3}}%
\smallskip\centerline{\vbox{\baselineskip12pt
\advance\hsize by -1truein\noindent\footnotefont{\bf Fig.~\the\figno:}
#2}}
\bigskip\endinsert\global\advance\figno by1}

\def\ifigure#1#2#3#4{
\midinsert
\vbox to #4truein{\ifx\figflag\figI
\vfil\centerline{\epsfysize=#4truein\epsfbox{#3}}\fi}
\narrower\narrower\noindent{\footnotefont
{\bf #1:}  #2\par}
\endinsert
}

\newwrite\ffile\global\newcount\tabno \global\tabno=1
\def\itab#1#2#3{\xdef#1{table~\the\tabno}
\goodbreak\midinsert\figin{\centerline{#3}}%
\smallskip\centerline{\vbox{\baselineskip12pt
\advance\hsize by -1truein\noindent\footnotefont{\bf Table~\the\tabno:}
#2}}
\bigskip\endinsert\global\advance\tabno by1}

\Title{\vbox{\baselineskip12pt\hbox{Univ.Roma I, 1146/96}
		\hbox{cond-mat/9701045}}}
{\vbox{\centerline{A Study of Activated Processes}
	\vskip2pt\centerline{in}
	\vskip2pt\centerline{Soft Sphere Glass}}}

\centerline{
David~Lancaster\footnote{$^{(a)}$}{\tt(lancaste@phys.uva.nl)} and
G.~Parisi\footnote{$^{(b)}$}{\tt(parisi@roma1.infn.it)} 
}

\bigskip
\centerline{{$^{(a)}$}Van der Waals -- Zeeman Lab.,}
\centerline{University of Amsterdam,}
\centerline{Valkenierstraat 65,}
\centerline{1018XE Amsterdam.}
\bigskip
\centerline{{$^{(b)}$}Dipartimento di Fisica and INFN,}
\centerline{Universit\`a di Roma I {\it La Sapienza},}
\centerline{Piazza A.~Moro 2, 00185 Roma.}

\vskip .3in
Abstract: 
On the basis of long simulations of a binary mixture of soft spheres
just below the glass transition, we make an exploratory study 
of the activated processes that contribute to the dynamics.
We concentrate on statistical measures of the size of the
activated processes.

%\draft
\Date{12/96} %replace this line by \draft  for preliminary versions
	     %or specify \draftmode at some point
\vfill\eject

%%%%%%%%%%%%%%%%%%%%%%%%%%%%%%%%%%%%%%%%%%%

\lref\GoAn{For reviews see, W. Gotze, {\it Liquid, freezing and the Glass
  transition}, Les Houches (1989), J.P.~Hansen, D.~Levesque,
  J.~Zinn-Justin editors, North Holland\semi 
  C.A.~Angell, Science, {\bf 267}, 1924 (1995).}
\lref\AdGi{G.~Adams and E.A.~Gibbs, J.Chem.Phys. {\bf 43},
  139 (1965).} 
\lref\Tglass{B.~Bernu, Y.~Hiwatari and J.P.~Hansen, 
  J.Phys. {\bf C18}, L371 (1985)\semi 
   B.~Bernu, J.P.~Hansen, Y.~Hiwatari and G.~Pastore,
  Phys.Rev. {\bf A36}, 4891 (1987)\semi 
   J.N.Roux, J.L.~Barrat and J.P.~Hansen,
  J.Phys. {\bf C1}, 7171 (1989).} 
\lref\KoAn{W.~Kob and H.C.~Andersen, 
  Phys.Rev. {\bf E51}, 4626 (1995)\semi 
  Phys.Rev. {\bf E52}, 4134 (1995).} 
\lref\MiHa{H.~Miyagawa, Y.~Hiwatari, B.~Bernu and J.P.~Hansen, 
  J.Chem.Phys {\bf 88}, 3879 (1988).} 
\lref\book{D.C.~Rapaport, {\it The Art of Molecular Dynamics Simulation},
  C.U.P.  (1995).} 
\lref\BaTe{See for example, C.~Battista {\it et al}, Int.J.High Speed 
  Comput. {\bf 5}, 637 (1993)\semi
  the use of APE for molecular dynamics simulations is also
  discussed by L.M.~Barone, R.~Simonazzi and A.~Tenenbaum, Roma I preprint.} 
\lref\Tfreeze{J.P.~Hansen, 
  Phys.Rev. {\bf A2}, 221 (1970)\semi
  W.G.~Hoover, M.~Ross, K.W.~Johnson, D.~Henderson, J.A.~Barker
   and B.C.~Brown, J.Chem.Phys. {\bf 52}, 4931 (1970).} 
\lref\Gioa{G.~Parisi, ``slow dynamics in glasses''.}

%%%%%%%%%%%%%%%%%%%%%%%%%%%%%%%%%%%%%%%%%%%%%%%%%%

\newsec{INTRODUCTION AND MOTIVATION}

It is well known that the relaxation times in glasses become 
extremely long when the glass transition is approached \GoAn. 
In fragile glasses, the relaxation times
definitely increase faster than a simple Arrenius behavior, 
for example the viscosity can be fitted by the Volger-Fulcher law, 
$\exp (A/(T-T_0))$.
In the standard picture of activated dynamics, 
this increase of the relaxation times implies an increase of the
energy barriers relevant to relaxation processes.
A divergence of the energy barriers at $T_0$ can hardly be explained 
without assuming some form of cooperative behaviour. 
Indeed, most theoretical proposals to explain
fragile glass behaviour assume that the dynamics is dominated by very
slow processes in which a large number of particles are rearranged \AdGi.

Extensive numerical simulations have been performed on glasses
\Tglass\KoAn, and the
behaviour of the diffusion constant (whose definition involves just 
a single particle) has been carefully studied. 
Unfortunately there are practically no studies of the relaxation
processes which occur in glasses and of their morphology 
(for example, the number of particles involved and their displacements). 
One of the most notable exceptions is the study of ref \MiHa\ where a 
rearrangement of 4 particles was observed.

The aim of the present study is essentially exploratory.
We simulate a binary mixture of soft spheres just below the 
glass temperature and address ourselves to the problem 
of identifying the activated processes and of studying their properties
in a systematic statistical way. 
We present the techniques we have used and the results we have obtained.
We have concentrated most of our attention on the
number of particles involved, or rather the spatial extent of the
activated process, and we present two different techniques to compute this
size.
A careful study of the temperature dependence of the quantities that 
we have measured would be extremely interesting, but it
goes beyond the limits of this work. Detailed theoretical
predictions for these quantities would also be welcome and
we hope that this note will stimulate research.

The paper is organized as follows: sections 2,3 and 4 describe the
simulation data and the remaining sections 5,6 and 7 are concerned 
with the analysis of activated processes.
In section 2 we discuss the model and the simulations that we have performed,
section 3 deals with the issue of thermalisation
and in section 4 we analyse our data using some of the distributions 
usually considered. 
In section 5 we consider the technique of cooling to
accurately find jumps and count them, sections 6 and 7 deal
with two methods of viewing the spatial distribution of the
displacements associated with the activated processes.
Because the study is exploratory, we give no conclusion.

%%%%%%%%%%%%%%%%%%%%%%%%%%%%%%%%%%%%%%%%%%%

\newsec{THE SIMULATIONS}

\subsec{The Model}

We use a Molecular Dynamics (MD) approach with leapfrog algorithm \book.
Because our goal is to look in detail at the 
activated processes underlying the dynamics, 
a small system is acceptable and we have used this to
advantage in writing a simple yet fast code running on
APE \BaTe\ by directly summing over all atom pairs rather than 
deal with the complications of neighbour lists that are
not simple on multi-processor machines.

We consider a total of $N=512$ spheres in a periodic box $8\times 8\times 8$, 
and use a standard technique to prevent the system crystalising
by working with a  50\% mixture of two different types of sphere 
with different effective radii \MiHa. The sphere species 
label is written $\alpha = 1,2$, the radii are
$\sigma_\alpha$ and the potential between spheres is,
\eqn\potential{
V_{\alpha \beta} = \left( {\sigma_{\alpha \beta} \over r}\right)^{12}
}
Where, $\sigma_{\alpha \beta}= (\sigma_\alpha + \sigma_\beta)/2$.
In this work we choose $\sigma_1= 1.0$ and $\sigma_2= 1.2$.
The masses of the spheres are both set equal to one.

This system of soft spheres and its variants have been studied extensively 
\refs{\Tglass,\MiHa} and the value of the melting and glass transition are 
known. The melting transition occurs at $T = 1.76\pm 0.06$
\Tfreeze\ while the glass transition is at $T = 0.56\pm 0.01$
\Tglass .
We shall work in the vicinity of $T=0.5$, just below the glass transition
where the dynamics is mainly due to the activated processes.
This is in contrast to the situation at higher temperature where smoother
mechanisms are responsible for diffusion. 
Indeed, the numerical method for finding the glass transition temperature is
as the temperature at which the smooth processes give vanishing diffusion.

\subsec{Data sets}

We have collected four data sets starting with 
different initial configurations; let us call them A,B,C,D. 
The individual characteristics of
each one, and the details of the initial state preparation are 
discussed in the appendix. Each data set is nominally at $T= 0.5$ and has the
same molecular dynamics parameters: that is a time step of 0.002
and a sampling of the configuration every $10^4$ such molecular
dynamics steps (corresponding to a time interval of 20 units). 
The total run time for each data set is always greater than $10^5$, 
but varies between data sets;
A: $2.6\times 10^5$, B: $5.2\times 10^5$, C: $1.4\times 10^5$, 
D: $1.8\times 10^5$.
A and B are the most interesting data sets, and for this reason, the longest. 
We find that the runs C and D are not really long enough to
obtain good statistics, but we have retained them for the purpose
of comparison.

\ifig\totalenergy{
Mean squared movement with respect to first configuration of the run,
small spheres (upper lines), and large spheres (lower lines).
Note that the time axis is scaled differently for the different length
runs.}
{\epsfysize=8cm \epsfbox{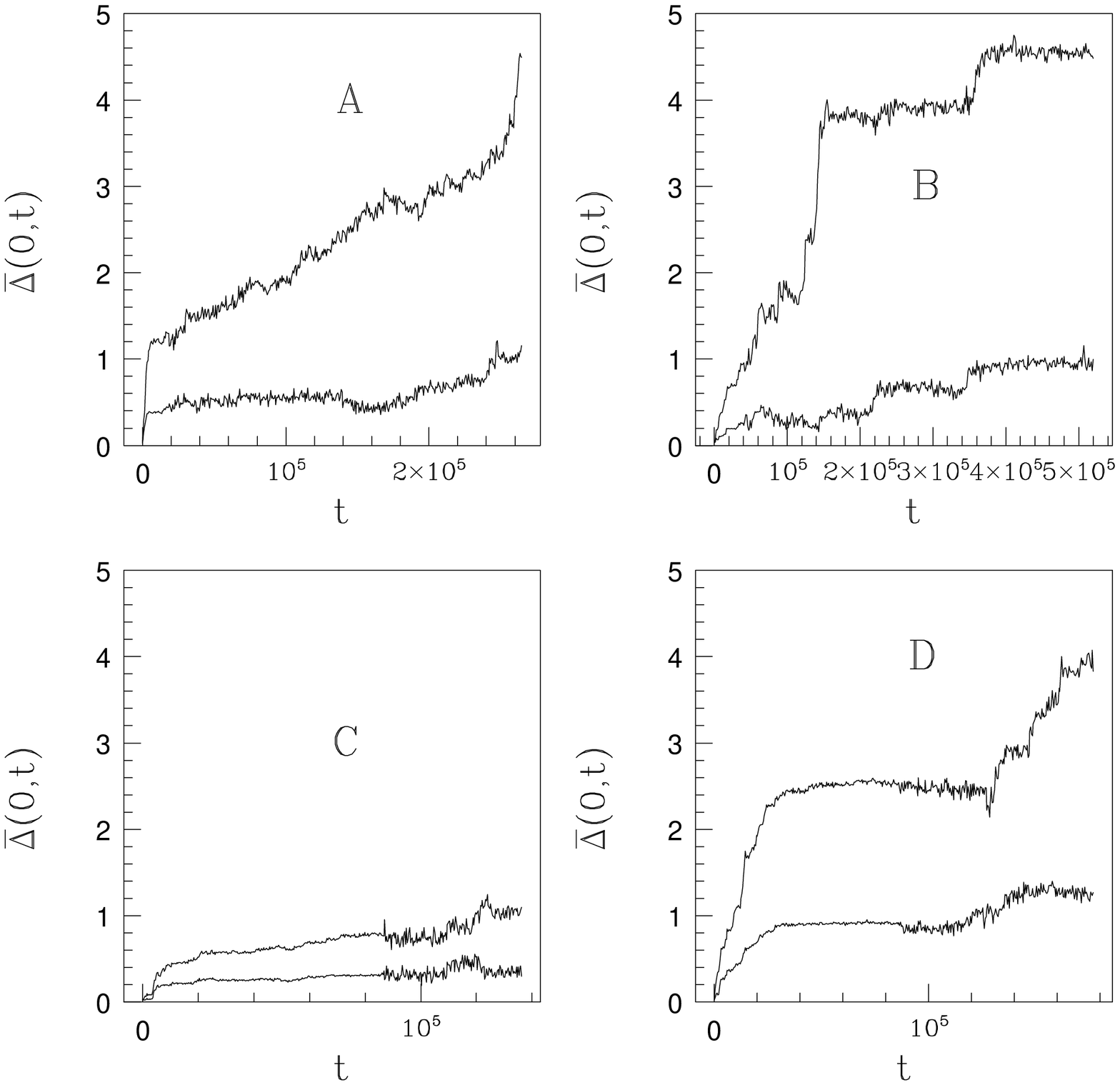}}

\subsec{Displacement}

The probability distribution for the distances moved by each
sphere after a given time gives a good initial idea of the
processes we are interested in analysing. 
The simplest quantity to consider is the mean squared distance
moved. When this is calculated in a periodic geometry there is
some latitude in the precise way of taking account
of spheres that have migrated around the periodic borders.
Here we have stored the configurations in such a way as
to keep track of this information by writing the coordinates,
$x \in (-\infty , \infty)$, as the periodic part,
$x_{pbc} \in (0,L)$,  plus the winding part that is an integral
multiple of $L$. 
We define the movement $\Delta_i(t,t')$ 
of the {\it i} th sphere between
times $t$ and $t'$ using coordinates with the centre of mass 
($\overline{\bf r} = {1\over N}\sum_i {\bf r}_i$) removed.
\eqn\deltarsq{
\Delta_i(t,t') 
= \bigl\vert {\bf r}_i(t) - {\bf r}_i(t')\bigr\vert^2
}
The mean squared movement is then,
\eqn\meandeltarsq{
\overline\Delta (t,t') 
= {1\over N} \sum_i \Delta_i(t,t')
}

In figure 1 we show the mean squared movement with respect to the
first configuration. There are wide variations in behaviour 
because the small temperature differences between the data sets have
a large effect on the rate of diffusion.
Note that in this, and subsequent figures the
time axis is scaled differently according to the length of
the different data sets.
This figure indicates that the movement is not
smooth, but progresses by a series of jumps which correspond
to the activated processes we are interested in.

%%%%%%%%%%%%%%%%%%%%%%%%%%%%%%%%%%%%%%%%%%%%%%%%%%

\newsec{THERMALISATION}

Because we work at temperatures slightly below the glass transition,
it is essential to discuss to what degree equilibrium is achieved
and to what extent our simulations are representative.

The molecular dynamics is at constant energy without 
any rescaling of momenta for stability during the course of
the simulations.
Our choice of MD time step is conservatively small, and 
the total energy always remained stable throughout the runs
to an accuracy of order $0.001$. 
No consistent drift was apparent even over the very longest 
time scales in data set B.

The data sets A and D
are at slightly higher total energy than the other pair and
each run is at a slightly different temperature.
The table below shows the  values of the energies averaged over
the duration of each complete run.
\itab\tabE{Average energies for each data set,
after removal of a thermalisation period of $0.5 \times 10^5$ 
(see discussion at the end of this section).} 
{$$\vbox{\offinterlineskip \halign{
\strut#&\vrule#\tabskip=1em plus 2em&
\hfil#\hfil& \vrule#&
\hfil#\hfil& \vrule#&
\hfil#\hfil& \vrule#&
\hfil#\hfil& \vrule#&
\hfil#\hfil& \vrule#\tabskip=0pt \cr
\noalign{\hrule}
& &\multispan{9}\hfil Data Set \hfil& \cr
\noalign{\hrule}
& & & & A& & B& & C& & D&  \cr
\noalign{\hrule}
& &Total E & &$7.326$& &$7.261$& &$7.254$& &$7.330$&  \cr
\noalign{\hrule}
& &Kinetic E & &$0.79$& &$0.77$& &$0.76$& &$0.80$&  \cr
\noalign{\hrule}
& &Potential E & &$6.54$& &$6.49$& &$6.50$& &$6.53$&  \cr
\noalign{\hrule}
}}
$$}

Data sets C and D are at slightly low and high temperatures respectively
and correspondingly we find few jumps in C and smoother behavior in D.

\ifig\kineticenergy{
Kinetic energy against time.
Each point has been averaged over 200 measurements
taken over a period of 800 time units (corresponding to
40 configurations).}
{\epsfysize=8cm \epsfbox{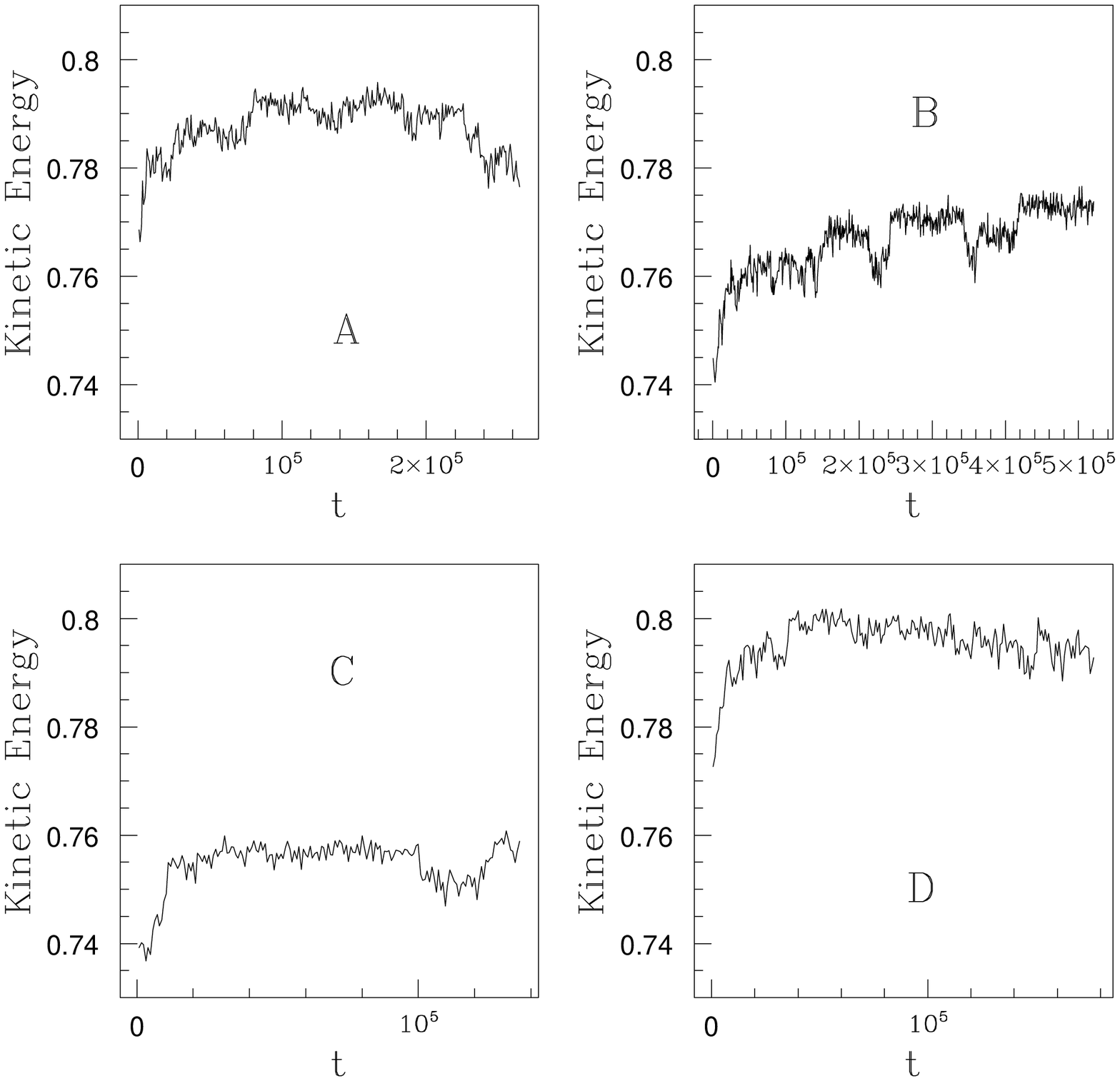}}

We have not given errors in the table because there is some consistent
relaxation during the course of the runs which becomes clear 
on a closer investigation of the kinetic energy shown in figure 2.
The potential energy decreases slightly and the kinetic part increases.
This is most noticeable in the beginning part of the plots
where it indicates insufficient thermalisation in the preparation of the
initial states. 
However, it is also clear that even over the very long time scales 
of data set B that relaxation continues.

These observations suggest that the
system relaxes to a lower potential well in the energy surface.
That this process can be seen in simulations is indicative of the smallness 
of the system and the very long runs.
With these data it is not possible to deduce anything concerning
dependence on the method of preparing the initial sample.

\ifig\rsqrelax{
Relaxation of average step movement with time.
Each point has been averaged over 100 measurements
taken over a period of 2000 time units.}
{\epsfysize=8cm \epsfbox{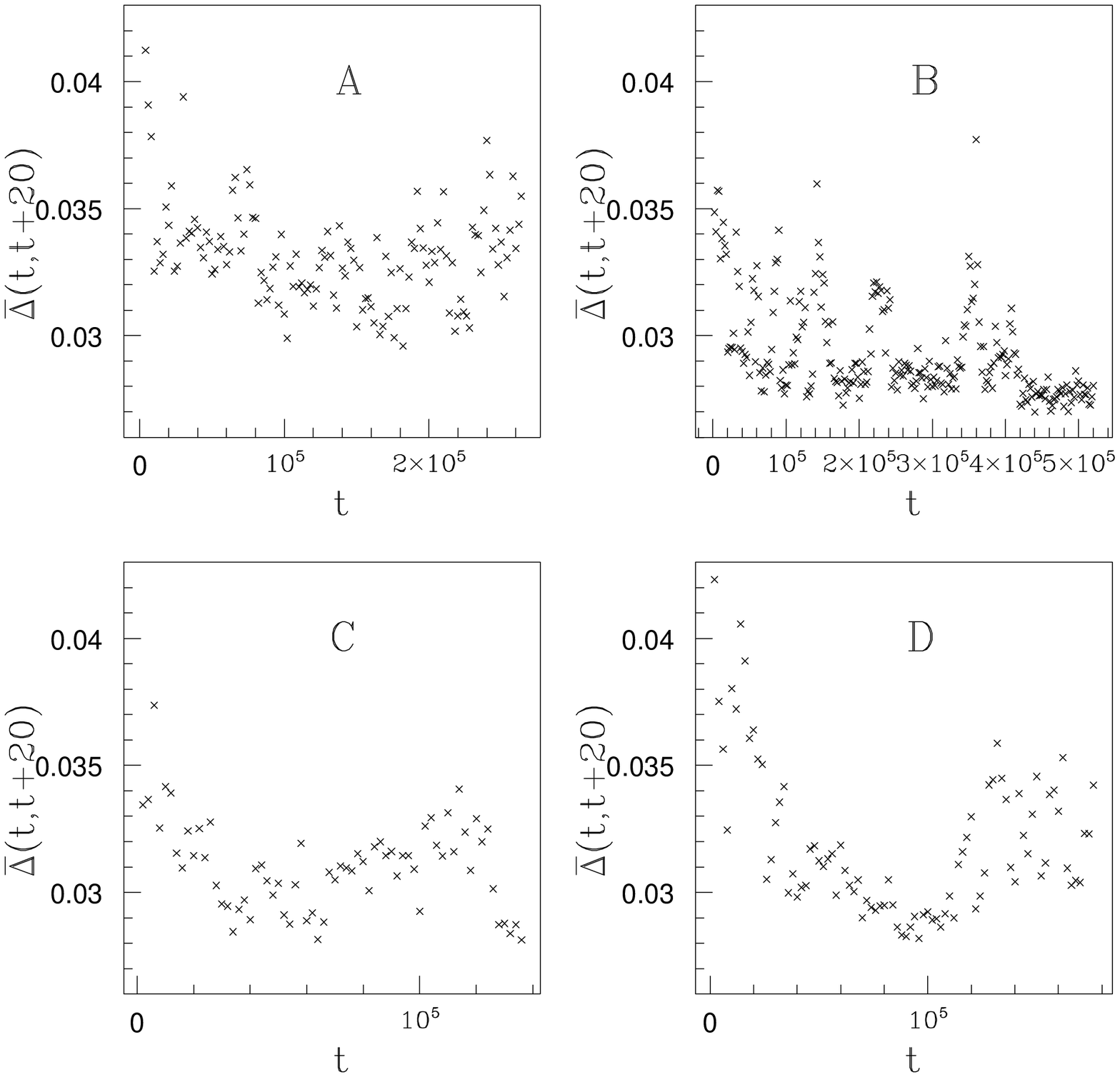}}
Another sign of the relaxation comes from the size of the move between
subsequent configurations. Since this data is relevant to later
developments we show it here. In \rsqrelax\ the value of 
$\Delta (t,t+20)$ between successive 
configurations is plotted. 
Each data point is averaged over 100 configurations.
Again an initial relaxation is apparent. It is noticeable that
even over the very long run B, there are significant variations
in the degree of activity. 

In subsequent sections where we consider time averaged
quantities we have removed an initial thermalisation period 
from the start of each data set.
The length of this thermalisation period has been fixed at
$0.5 \times 10^5$ time units from a study of \kineticenergy\
and \rsqrelax .
We see that this procedure still leaves some long time relaxation 
and variation in degree of activity in the case of B.
This observation throws some light on the puzzling behaviour of the
lowest temperature data set, C, which shows a suprisingly high degree of 
activity in \rsqrelax\ despite its small overall movement
shown in \totalenergy. It seems likely that the whole of the
run C corresponds to one of the active periods of B and that a thermalisation
time longer than  $0.5 \times 10^5$, and in fact longer than the whole run,
is needed.
This should be bourne in mind for later analyses.

%%%%%%%%%%%%%%%%%%%%%%%%%%%%%%%%%%%%%%%%%%%%%%%%%%

\newsec{DIFFUSION DISTRIBUTIONS}

Various probability distributions of the
step movement have been investigated at higher temperatures in the
work of Hansen {\it et al.} \Tglass\ for soft spheres
and more recently for Leonard Jones spheres by Kob and Andersen \KoAn. 
We first consider the probability distribution of the movements
of the individual spheres between configurations, 
$P_1\bigl(\Delta(t,t+20)\bigr)$.
This is defined as,
\eqn\Pdist{
P_1(\Delta) =
{1\over N}\sum_i \delta\bigl(\Delta - \Delta_i(t,t+20)\bigr)
}
When calculated for a single pair of configurations, 
the distribution 
is noisy but the tail corresponding to activated processes is
visible in cases where there is a jump. 
\ifig\Pofrsq{
Averaged probability distribution of step movement, $P_1(\Delta)$, 
shown with a 
logarithmic scale. Upper and lower curves are for small and large
spheres respectively.}
{\epsfysize=8cm \epsfbox{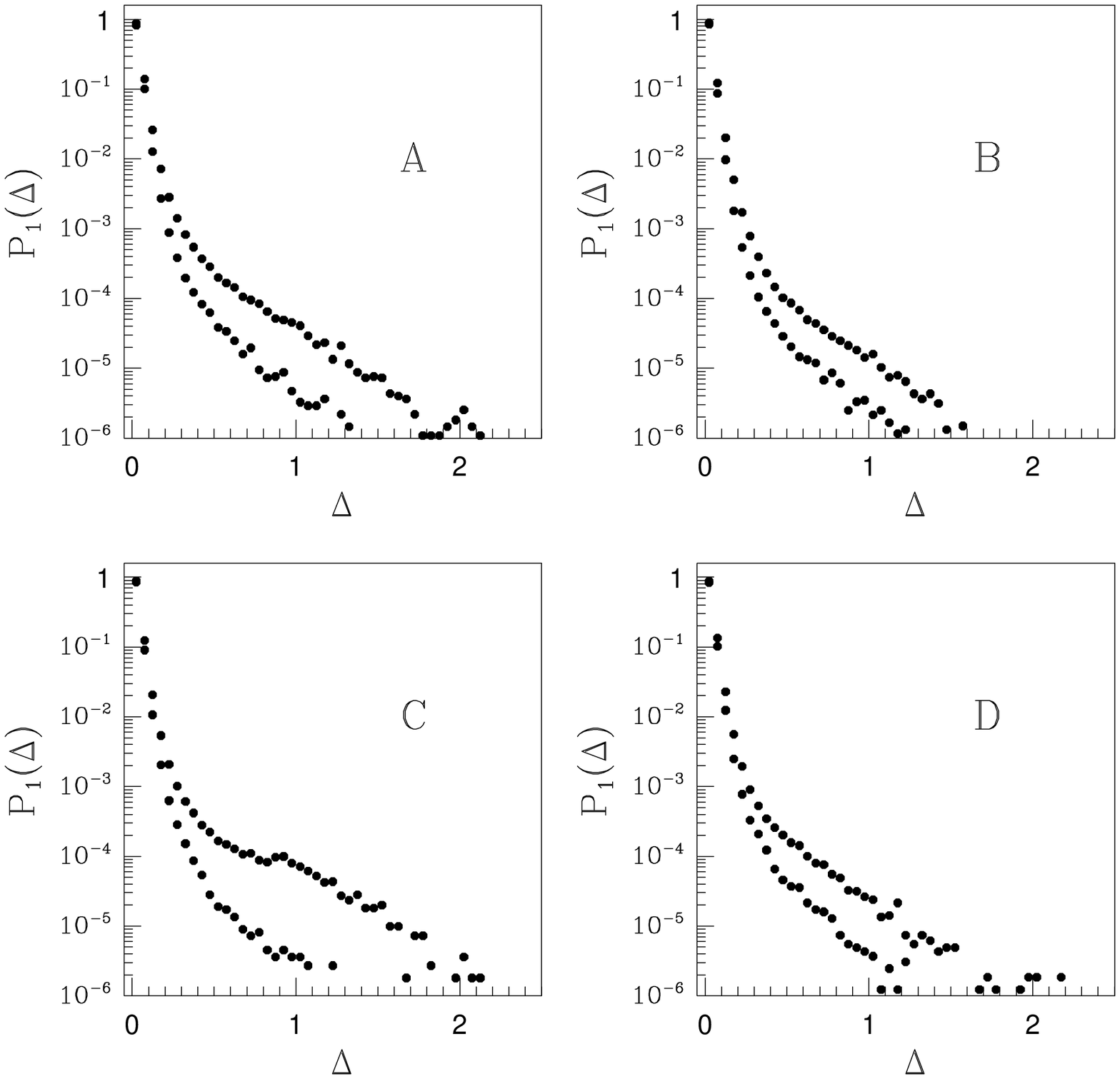}}
In figure 4 we show
this quantity averaged over the time duration of the data set
(with initial thermalisation period removed).
We use a logarithmic scale to see the small contribution of large
motions.

The tail of the distribution shows behavior indicative of 
diffusion since the probability distribution decays exponentially 
with $\Delta$. The diffusion is due to the activated processes,
in contrast to the situation at higher temperature where smoother
mechanisms are responsible. The time step of 20 units is not long
enough to justify the distribution one would expect from diffusion, 
$\sim \exp (-\Delta /4Dt)$. If nevertheless we use this formula to
define an effective diffusion constant, we find 
values of order, $3\sim 4\times 10^{-3}$, which are somewhat
larger than the values found by other methods in \Tglass. 
\ifig\Pofavrsq{
Probability distribution of mean squared step movement,
$P_2(\Delta)$.
Shown only for small spheres.}
{\epsfysize=8cm \epsfbox{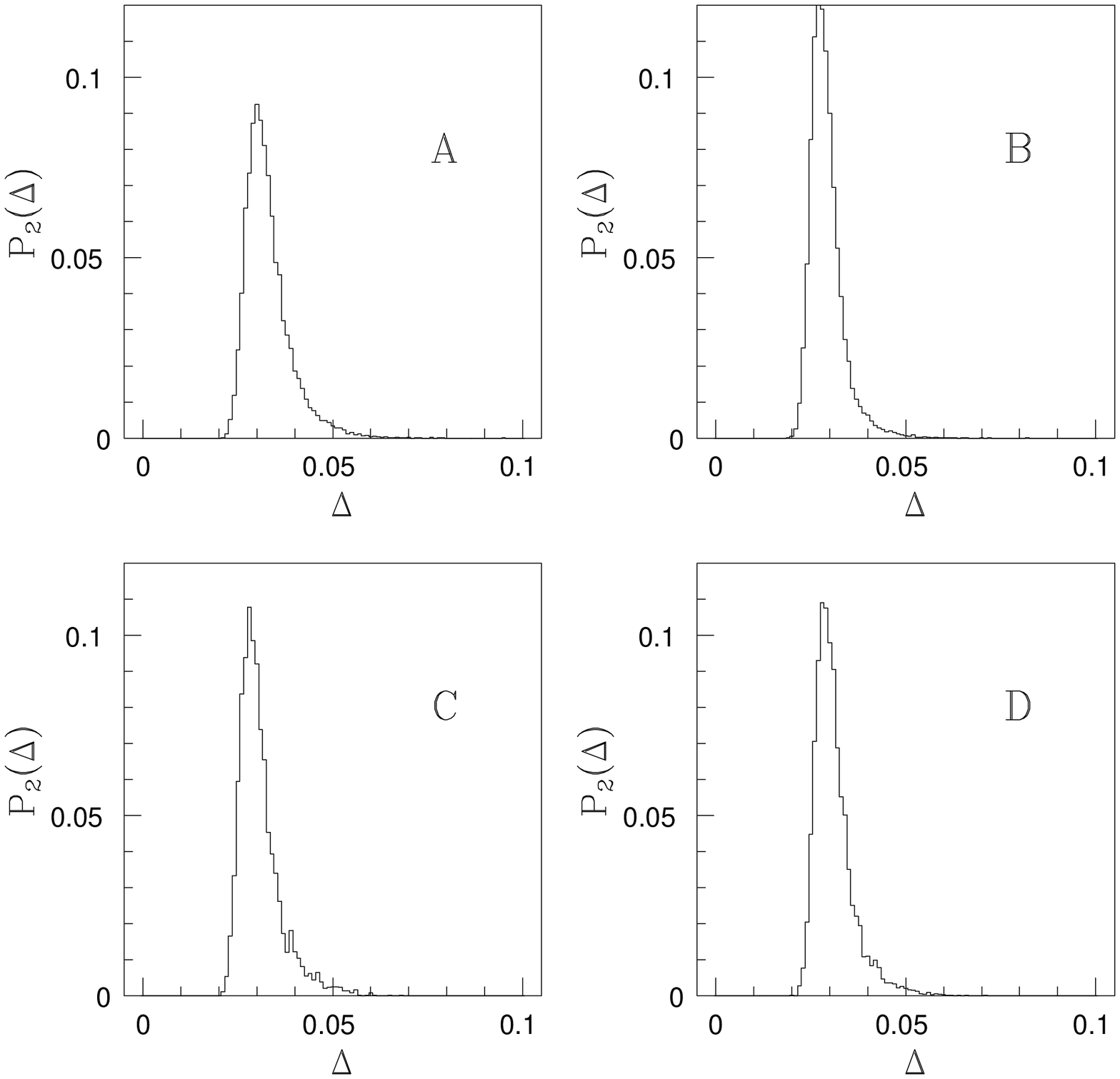}}

It is sometimes useful to consider probability distributions in time.
In \Pofavrsq\ we show the probability distribution
of the mean square step displacement, $P_2({\overline\Delta}(t,t+20))$. 
For a run of $S$ configurations this is defined as,
\eqn\Pmeandist{
P_2(\Delta) =
{1\over S}\sum_t^S \delta\bigl(\Delta - {\overline\Delta(t,t+20)}\bigr)
}
In comparison with \Pdist, this probability distribution
shows the variations in time of a spatially averaged movement.
The tail should indicate the likelihood of jumps, 
but there is not sufficient data to determine its behavior accurately.
One would expect a tall narrow peak with small tail at low temperature,
and a somewhat wider peak shifted to larger $\overline \Delta$ at higher
temperature. Comparison of A and B correctly identifies their
temperature ordering, but there is insufficient data to make
similar statements with regard to the shorter runs C and D.

%%%%%%%%%%%%%%%%%%%%%%%%%%%%%%%%%%%%%%%%%%%%%%%%%%

\newsec{COOLING AND LOCATING JUMPS}

\subsec{Cooling}

We have already seen that the motion of the system does not
proceed smoothly, but rather by a series of jumps. We now turn our
attention to a detailed analysis of these activated processes.
The most clear definition of an activated process would be a movement 
between different potential wells.  This can only be determined
unambiguously in the absence of thermal fluctuations and motivates 
us to study configurations that have been cooled to zero temperature.
\ifig\hotvscool{
Comparison of hot and cooled data for a fairly active period
of $0.2\times 10^5$ time units taken from data set B.
Mean squared displacement, $\overline\Delta$, is shown
with respect to the first configuration of this time interval.}
{\epsfysize=6cm \epsfbox{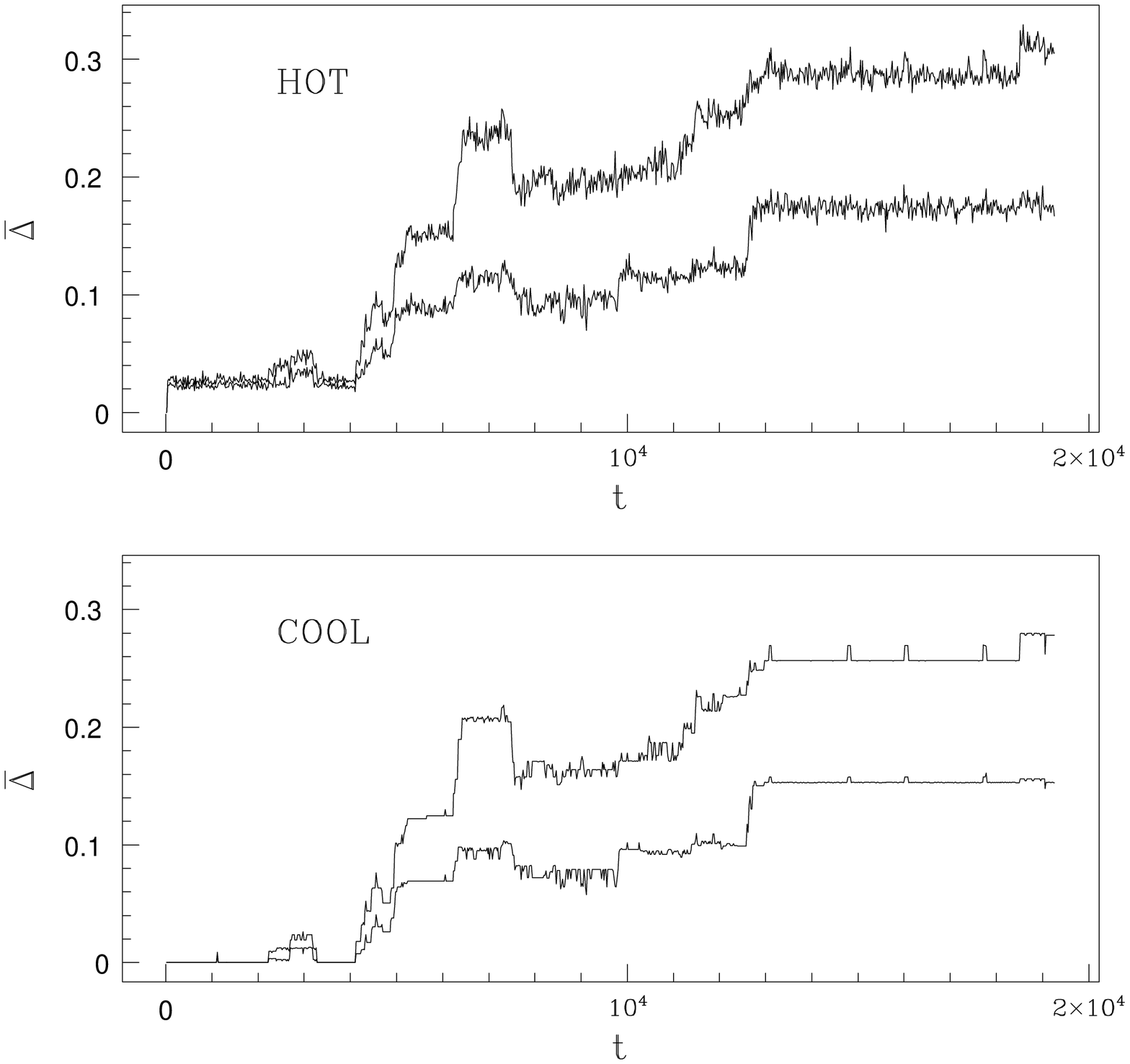}}

The method of cooling we chose was to introduce an effective friction 
into the MD equations by reducing the momenta by a factor of 0.998
each time step. This technique was found to be more efficient than
a steepest descent type method in which the initial momenta are
ignored. We made 5000 MD steps to cool each configuration
and typically reduced the kinetic energy to order $10^{-5}$.
We have not cooled all configurations
obtained, but have decided to look at a fairly active part of the
time evolution of data set B consisting of a period of
$0.2\times 10^5$ time units fairly early in the complete run.
We have cooled the 1000 configurations from 3854 to 4854 of data set B. 
The effect of cooling is clearest if we consider the total movement in the
cooled set and compare with the original hot data.
Inspection of \hotvscool\ makes it clear how thermal noise is removed.

In this time period we find 261 jumps which can be classified
as one of two types.
Some of the jumps merely interchange a small
number of particles and simply amount to a relabeling of the
configuration, whereas others are less simple and change the energy. 
For example, in the first class we have seen motions which only 
relabel between 2 and 5 particles, with all 
other particles remaining fixed.
In the other class the major part of the movement is still local,
but all other particles must move by a small amount in
order to accommodate the new configuration.
In this case it is less easy to say how many particles are 
involved in the jump. 
The potential energy change occurring in the second class of
jump is sometimes very small and only related to a metastable
state. In these circumstances the cooling is providing excessive 
information about small features of the potential surface
which would be washed out in a finite temperature simulation.
This effect is also apparent in the cool plot of \hotvscool\ 
where many of the 261 jumps are small and should not be
regarded as important processes.

\itab\tabdeltaHC{Fraction of jumps with size 
(according to various criteria) greater than some cutoff. Shown for 
the set of 1000 cooled configurations and their hot counterparts.}
{$$\vbox{\offinterlineskip \halign{
\strut#&\vrule#\tabskip=1em plus 2em&
\hfil#\hfil& \vrule#&
\hfil#\hfil& \vrule#&
\hfil#\hfil& \vrule#\tabskip=0pt \cr
\noalign{\hrule}
& & & & COOL& & HOT&  \cr
\noalign{\hrule}
& &All jumps & &$0.271$& &$\sim$&  \cr
\noalign{\hrule}
\noalign{\hrule}
& &${\overline\Delta}>0.01$ & &$0.089$& &$1.0$&  \cr
\noalign{\hrule}
& &${\overline\Delta}>0.02$ & &$0.017$& &$1.0$&  \cr
\noalign{\hrule}
& &${\overline\Delta}>0.03$ & &$0.002$& &$0.411$&  \cr
\noalign{\hrule}
& &${\overline\Delta}>0.04$ & &$0.0$& &$0.074$&  \cr
\noalign{\hrule}
\noalign{\hrule}
& &$\overline{\Delta^2}>0.002$ & &$0.133$& &$0.368$&  \cr
\noalign{\hrule}
& &$\overline{\Delta^2}>0.003$ & &$0.091$& &$0.201$&  \cr
\noalign{\hrule}
& &$\overline{\Delta^2}>0.004$ & &$0.069$& &$0.154$&  \cr
\noalign{\hrule}
& &$\overline{\Delta^2}>0.005$ & &$0.060$& &$0.117$&  \cr
\noalign{\hrule}
\noalign{\hrule}
& &$\Delta_{max}>0.3$ & &$0.184$& &$0.300$&   \cr
\noalign{\hrule}
& &$\Delta_{max}>0.4$ & &$0.156$& &$0.209$&   \cr
\noalign{\hrule}
& &$\Delta_{max}>0.5$ & &$0.128$& &$0.163$&   \cr
\noalign{\hrule}
& &$\Delta_{max}>0.6$ & &$0.086$& &$0.128$&   \cr
\noalign{\hrule}
}}$$}

Given this technique of unambiguously finding jumps in the cooled data
it is interesting to see whether their presence can be accurately
predicted by a study of the hot data. A direct procedure would be to 
introduce a cutoff on the displacement. We find that this certainly captures 
the larger jumps in the cooled data, but not all the smaller ones.
On the other hand, as mentioned above, small moves found by the method of
cooling need not have any significance.
A good cutoff is a matter of empirical choice; in table 2 
we consider cutoffs in the mean squared displacement
$\overline \Delta (t,t+20)$, the mean quartic displacement
$\overline {\Delta^2}(t,t+20)$ and  the maximum (amongst
the spheres) step size $\Delta_{max}(t,t+20)$. 
A cutoff on the energy change is not effective.
We use the same cutoff for large and small species of sphere,
but in effect it is always the small spheres that signal a jump.
We show the fraction of jumps found in this cooled set of configurations
calculated using each criterion.
The thermal motion makes a cutoff in $\overline\Delta$ insensitive
and we shall discard this method.
Of the other choices, the cutoff on $\Delta_{max}$
seems to give closer results between the hot and cool data 
so we prefer this technique.

\itab\tabdeltaAD{Fraction of jumps with size 
($\Delta_{max}$) greater than some cutoff. Shown for 
complete thermalised data sets.}
{$$\vbox{\offinterlineskip \halign{
\strut#&\vrule#\tabskip=1em plus 2em&
\hfil#\hfil& \vrule#&
\hfil#\hfil& \vrule#&
\hfil#\hfil& \vrule#&
\hfil#\hfil& \vrule#&
\hfil#\hfil& \vrule#\tabskip=0pt \cr
\noalign{\hrule}
& &\multispan{9}\hfil Data Set \hfil& \cr
\noalign{\hrule}
& & & & A& & B& & C& & D&  \cr
\noalign{\hrule}
\noalign{\hrule}
& &$\Delta_{max}>0.3$ & &$0.374$& &$0.192$& &$0.407$& &$0.264$&  \cr
\noalign{\hrule}
& &$\Delta_{max}>0.4$ & &$0.235$& &$0.105$& &$0.212$& &$0.171$&  \cr
\noalign{\hrule}
& &$\Delta_{max}>0.5$ & &$0.165$& &$0.071$& &$0.146$& &$0.122$&  \cr
\noalign{\hrule}
& &$\Delta_{max}>0.6$ & &$0.119$& &$0.050$& &$0.113$& &$0.084$&  \cr
\noalign{\hrule}
}}
$$}

Having identified suitable ranges for the cutoffs
we use the same method to analyse the complete thermalised data sets.
Table 3 shows the fraction of jumps observed
throughout the runs. 

From a comparison of the tables it is clear that the set of 
configurations we cooled were indeed more active than the average 
of data set B.
In table 3, A has a larger fraction of jumps than B, as we would
expect since it is slightly warmer. The shorter runs, C and D,
do not however fit this pattern, presumably because of their limited data.

In a later section we will use this method to identify jumps
and will fix the cutoff at the  conservative end of the range,
$\Delta_{max} > 0.6$, in order to eliminate spurious processes 
relating to metastable states.

%%%%%%%%%%%%%%%%%%%%%%%%%%%%%%%%%%%%%%%%%%%%%%%%%%

\subsec{Viewing Activated Processes}

Activated processes are local disturbances and
an intuitive way of visualising them is to locate the centre
and plot the radial variation in the move size.

We choose to define the centre as the mean position weighted by
the displacement. 
\eqn\centre{
{\bf r}_c = 
{\sum_i {\bf r} (\Delta_i)^\alpha
\over
\sum_i (\Delta_i)^\alpha}}
It does not correspond to the location of any particle.
Because of the periodic boundary conditions this definition
actually requires some prior guess which is 
obtained from the location of the largest move.
For cooled data the method is then straightforward, but for hot
data another complication arises.
Random thermal movements of spheres distant from the true centre
contribute excessively to the mean and must be suppressed.
We do this by weighting with a power, $\alpha$, of the movement, rather
than the movement itself. Empirically, a power of $\alpha = 2$ 
(corresponding to weights $\Delta^2$) is found to be adequate.
\ifig\viewjump{
Example of the movement of spheres plotted against the distance 
from the centre of the jump.
Hot (left) and cooled (right) data, and different weight
factors $\Delta$ (top) and  $\Delta^2$ (bottom).
The example comes from configuration 4075 to the subsequent
configuration of data set B.
The location of the centre varies slightly in each plot.}
{\epsfysize=8cm \epsfbox{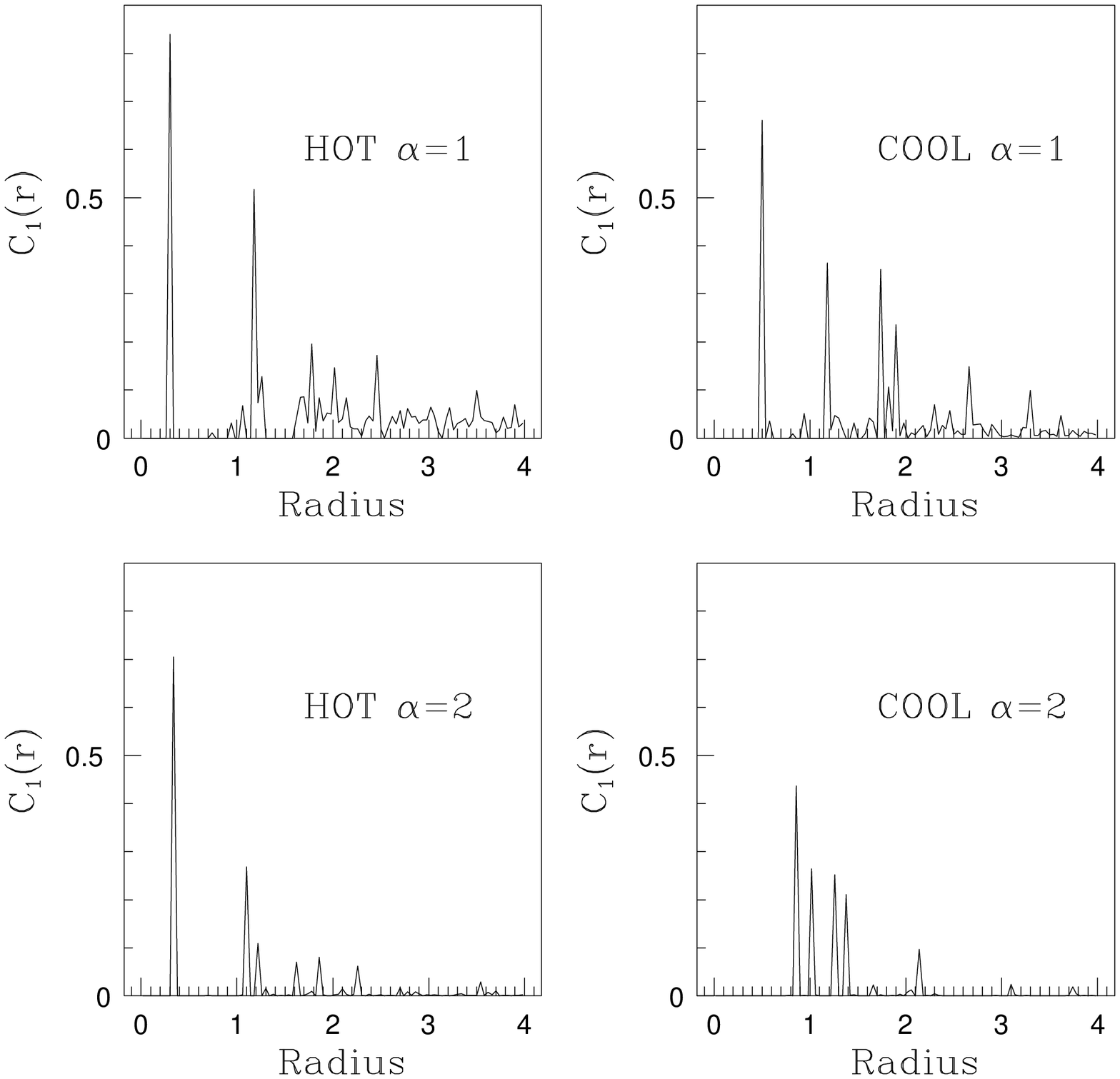}}
Once the centre is defined, the radial distribution is given by,
\eqn\corrone{
C_1(r) =
{\sum_i \Delta_i^\alpha \delta \bigl(
r - |{\bf r}_i - {\bf r}_c| \bigr)
\over
\sum_i \delta \bigl(
r - |{\bf r}_i - {\bf r}_c| \bigr)}
}
Figure 7 shows an example of this distribution for a particular
jump involving quite a large number of particles.
The plot is repeated with both hot and cooled data and also using  
$\Delta$ and  $\Delta^2$  in the weighting factor.
The effectiveness of the weighting factor in reducing
thermal noise at large radius is apparent.
\ifig\moveszdist{
Averaged movement of spheres, $C_1(r)$, plotted against the distance 
from the centre of the jump.
Using hot data with a cutoff, $\Delta_{max} > 0.6$ on
all thermalised data. We have used the weighting power $\alpha = 2$.}
{\epsfysize=8cm \epsfbox{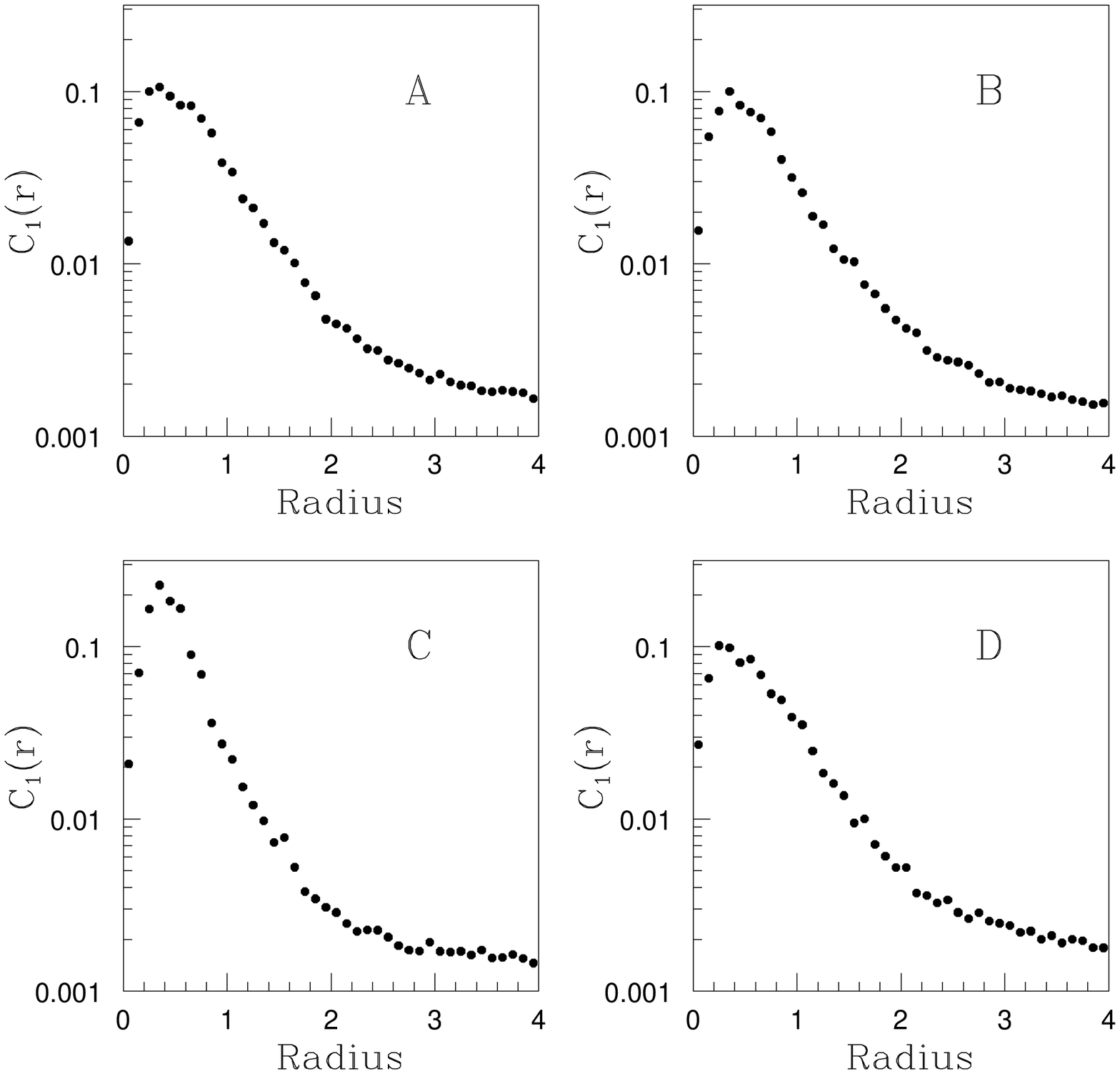}}

In applying the method to hot data it is important
that a jump is truly present, otherwise the centre 
will correspond to the location of 
some slightly larger than average random movement. 
To check that there is a jump, a cutoff on the maximum move 
can be introduced as discussed in the previous section.
In figure 8 we use this technique to show an
averaged form of the distribution $C_1(r)$. 
We impose the conservative requirement for a jump by taking
the average only over jumps characterised by 
$\Delta_{max} > 0.6$ in the thermalised data. 
We have plotted the distribution using a logarithmic
scale to bring out the small contributions at large 
radius from the centre.
The variation between runs of the mean size for the jumps
is too small  to be able to analyse temperature dependence.

In summary; this technique is helpful for intuition in
seeing individual jumps but relies on too many parameters to
be a good statistical measure for activated processes.

%%%%%%%%%%%%%%%%%%%%%%%%%%%%%%%%%%%%%%%%%%%%%%%%%%

\newsec{CORRELATION}

In this section we present an
alternative method of determining the size of movement without the
need to prejudge the presence of a jump by introducing a cutoff.
We evaluate correlations between the displacements of
different spheres by calculating the following,
\eqn\corrln{
C_2(r)=
{\sum_{ij,|{\bf r}_i- {\bf r}_j| < R}
 \left(\Delta_i(t,t+20) - \overline\Delta\right)
 \left(\Delta_j(t,t+20) - \overline\Delta\right)
\delta(r - \vert{\bf r}_i - {\bf r}_j\vert)
\over
\sum_{ij,|{\bf r}_i- {\bf r}_j| < R}
\delta(r - \vert{\bf r}_i - {\bf r}_j\vert)}
}
The quantity in the denominator is well known as the structure function.
In the numerator, note that we have subtracted the mean value of the 
displacement.
This avoids the difficulties experienced in the previous
method of requiring weighting by powers of $\Delta$
in order to suppress long distance thermal motion.
$R$ is the maximum distance on the periodic volume, which is 4.0 in our case.
\ifig\correlation{
Correlations, $C_2(r)$, averaged over thermalised data.
An offset of $2\times 10^{-5}$ has been added to allow 
a logarithmic axis.}
{\epsfysize=8cm \epsfbox{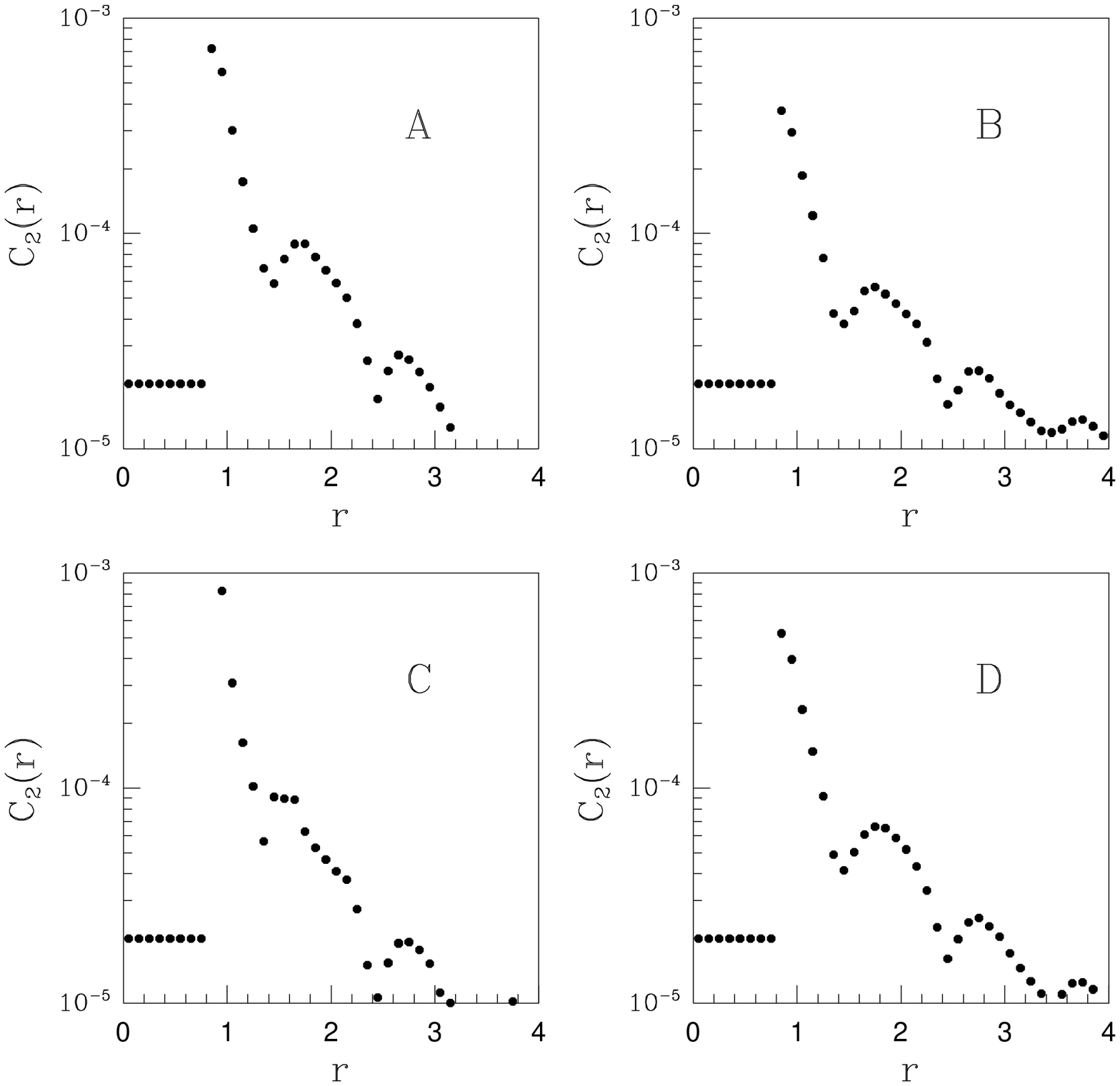}}

The quantity $C_2(r)$ 
contains little information when evaluated at a single time for a pair of
configurations. But when averaged over the runs as shown
in \correlation\ it displays interesting structure.
We see some oscillation as we expect for a quantity similar to the
structure function, but the major contribution is from adjacent
spheres. Since $C_2(r)$ tends to go slightly negative
at large $r$ we have added a constant offset in order to be able to use
a log scale. This offset is clear in the figure since it 
corresponds to the plateau at small $r$ where $C_2(r)$ strictly vanishes.
The mean jump size, $\int C(r) r dr$, varies little between runs.

It would be interesting to have theoretical predictions and
better measurements for the temperature dependence of this
correlator.

\bigbreak\bigskip\bigskip\centerline{{\bf Acknowledgements}}\nobreak
DL would like to thank the British Council and the Consiglio
Nazionale delle Ricerche for financial support during part of this work.

\appendix{}{Initial States of A,B,C,D.}

In this appendix we discuss the individual characteristics of
the data sets A,B,C,D, and the details of the initial state preparation.

All the initial configurations derive from a high temperature
molecular dynamics run sampled at time intervals that yield 
independent configurations.
To be precise, the high temperature is $T=8.0$ where we 
use a version of molecular dynamics 
that renews momenta from a gaussian distribution 
every 5000 steps with time step 0.001.
The initial thermalisation is of 3 million steps (corresponding to
a total movement per particle of 
${\overline \Delta } \sim 2\times 10^4$), 
then four configurations
are taken at intervals of $5\times 10^5$ steps (which corresponds to
a movement per particle of 
${\overline \Delta} \sim 3.6\times 10^3$).
These four independent configurations are then treated as follows.

A) An slow annealing of 3 million steps down to $T=0.5$. The last
configuration of the annealing is the initial configuration for 
data set A. 

B) A more abrupt annealing of $7.5\times 10^5$ steps down to $T=0.5$
followed by a pre-thermalisation consisting of $4\times 10^5$ steps
with momentum updates intended to stabilise the temperature.
It is some of the later configurations of this data set, B,
that have been cooled as discussed in section 5.

C) A direct quench down to $T=0.5$ by rescaling the momenta
followed by a brief pre-thermalisation consisting of 2000 steps.
No jumps of the type in figure 1 are observed in this data set 
which is at a slightly lower
temperature, and it has therefore not been extended as far as A or B.

D) A history like that of set B, with an annealing followed
by pre-thermalisation with the same parameters as B.
In this case the temperature turns out to be slightly higher
than in the other runs and for this reason the jumps we observe 
are not very sharp.

\listrefs
%\listfigs
\bye